\documentclass[conference]{IEEEtran}
\IEEEoverridecommandlockouts
\usepackage{cite}
\usepackage{amsmath,amssymb,amsfonts}
\usepackage{algorithm}

\usepackage{textcomp}
\def\BibTeX{{\rm B\kern-.05em{\sc i\kern-.025em b}\kern-.08em
    T\kern-.1667em\lower.7ex\hbox{E}\kern-.125emX}}

\usepackage[pdftex]{graphicx}

\usepackage{amsmath}

\usepackage[noend]{algpseudocode}
\usepackage{array}
\ifCLASSOPTIONcompsoc
 \usepackage[caption=false,font=normalsize,labelfont=sf,textfont=sf]{subfig}
\else
 \usepackage[caption=false,font=footnotesize]{subfig}
\fi

\usepackage{stfloats}
\usepackage{url}
\usepackage{amssymb} % checkmark
\usepackage{multirow}
\usepackage{hyperref}  %hyperref still needs to be put at the end!
\usepackage[usenames,dvipsnames,svgnames]{xcolor}
\usepackage{colortbl}
\usepackage{hhline}

\usepackage{pgfplots}
\pgfplotsset{compat=1.18}

\makeatletter
\let\old@ps@headings\ps@headings
\let\old@ps@IEEEtitlepagestyle\ps@IEEEtitlepagestyle
\def\confheader#1{%
\def\ps@IEEEtitlepagestyle{%
\old@ps@IEEEtitlepagestyle%
\def\@oddhead{\strut\hfill#1\hfill\strut}%
\def\@evenhead{\strut\hfill#1\hfill\strut}%
}%
\ps@headings%
}
\makeatother

\confheader{%
\textit{To appear in the 33rd International Conference on Field-Programmable Logic and Applications (2023).}
}

\usepackage[pscoord]{eso-pic}
\newcommand{\placetextbox}[3]{
\setbox0=\hbox{#3}
\AddToShipoutPictureFG*{ \put(\LenToUnit{#1\paperwidth},\LenToUnit{#2\paperheight}){\vtop{{\null}\makebox[0pt][c]{#3}}}
}
}
\placetextbox{.5}{0.055}{\small \copyright 2023 IEEE. Personal use of this material is permitted. Permission from IEEE must be obtained for all other uses, in any current or}
\placetextbox{.5}{0.045}{\small future media, including reprinting/republishing this material for advertising or promotional purposes, creating new collective works,}
\placetextbox{.5}{0.035}{\small for resale or redistribution to servers or lists, or reuse of any copyrighted component of this work in other works.}

\begin{document}

\title{Compiler Discovered Dynamic Scheduling of Irregular Code in High-Level Synthesis}

% \author{\IEEEauthorblockN{Anonymized}
% \IEEEauthorblockA{\textit{Anonymized}\\
% Anonymized \\
% Anonymized}
% }
\author{\IEEEauthorblockN{Robert Szafarczyk, Syed Waqar Nabi and Wim Vanderbauwhede}
\IEEEauthorblockA{School of Computing Science\\
University of Glasgow, UK
\\ Email: \{robert.szafarczyk, syed.nabi, wim.vanderbauwhede\}@glasgow.ac.uk}}

\maketitle

\begin{abstract}

Dynamically scheduled high-level synthesis (HLS) achieves higher throughput than static HLS for codes with unpredictable memory accesses and control flow.
However, excessive dataflow scheduling results in circuits that use more resources and have a slower critical path, even when only a part of the circuit exhibits dynamic behavior.
Recent work has shown that marking parts of a dataflow circuit for static scheduling can save resources and improve performance (hybrid scheduling), but the dynamic part of the circuit still bottlenecks the critical path.

We propose instead to selectively introduce dynamic scheduling into static HLS.
This paper presents an algorithm for identifying code regions amenable to dynamic scheduling and shows a methodology for introducing dynamically scheduled basic blocks, loops, and memory operations into static HLS. 
Our algorithm is informed by modulo-scheduling and can be integrated into any modulo-scheduled HLS tool.
On a set of ten benchmarks, we show that our approach achieves on average an up to 3.7$\times$ and 3$\times$ speedup against dynamic and hybrid scheduling, respectively, with an area overhead of 1.3$\times$ and frequency degradation of 0.74$\times$ when compared to static HLS.
% using fewer resources and achieving lower critical paths.

\end{abstract}


\section{Introduction}

High-level synthesis (HLS) tools transform code written in a high-level software language, like C\texttt{++}, into a hardware description of a custom architecture that can be realized on FPGAs. 
Custom architectures can achieve a higher degree of pipeline parallelism compared to superscalar CPUs and GPUs, promising performance and efficiency improvements \cite{efficient_communication_analysis_adler}.
This performance promise has led to wider adoption of FPGA acceleration \cite{amazon_f1_fpga, microsoft_fpga}.
The success of such acceleration depends in part on the quality of HLS tools.
% and the slowdown in the increase of CPU speeds,

A major objective of HLS tools is loop pipelining.
Loop pipelining is the process of starting new iterations of a loop while previous iterations have not yet finished.
The number of cycles between the start of two iterations is called the Initiation Interval (II).
A loop with a constant II, $N$ iterations, and a latency of $L$ will execute in $L + (N - 1) \times II$ cycles, which for $N \gg L$ can be approximated as $N \times II$.
Thus, a low loop II is crucial to achieving good performance in HLS.

\begin{figure*}[t!]
\centering
\begin{minipage}[b]{.25\textwidth}
% \vspace{-3em}
\subfloat[Motivating source code.]{{\includegraphics[width=1\textwidth]{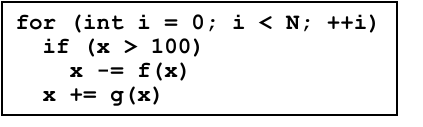}}%
\label{fig:MotivatingExampleCode}}
\vfill
\subfloat[Data and control dependence graph.]{{\includegraphics[width=1\textwidth]{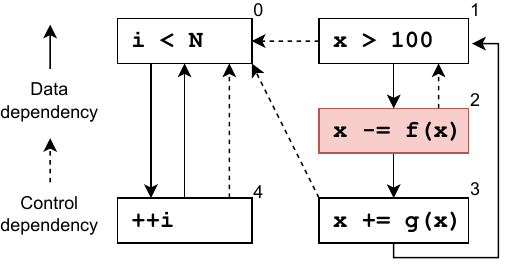}}%
\label{fig:MotivatingExampleDDG}}
\end{minipage}%
\hfill%
\begin{minipage}[b]{.74\textwidth}
\subfloat[A static schedule: a new iteration started every 5 cycles for all iterations.]{\includegraphics[width=1\textwidth]{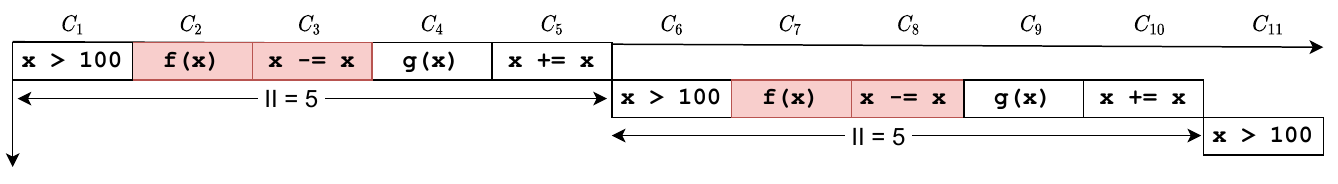}%
\label{fig:MotivatingExampleStaticSchedule}}
\vfill
\subfloat[An ideal schedule: \texttt{x = x - f(x)} is never allocated a slot if not required.]{\includegraphics[width=1\textwidth]{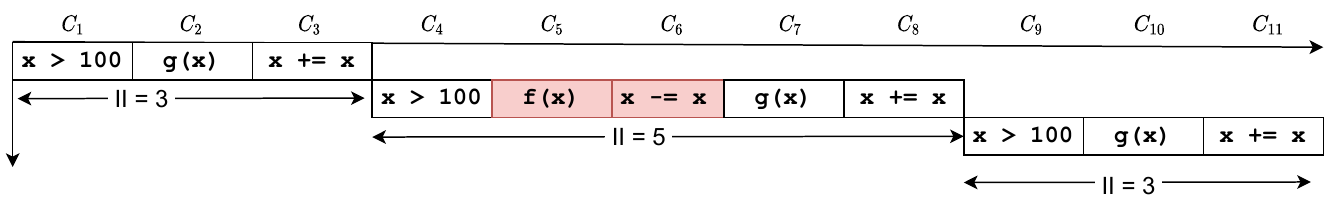}%
\label{fig:MotivatingExampleIdealSchedule}}

\end{minipage}

\caption{A motivating example of code with an inter-iteration control-dependent data data dependency (a). Current modulo-scheduled HLS needs to create a worst case schedule (c). We propose to enhance modulo-scheduled HLS with analysis and transformation passes which enable the dynamic schedule in (d).}

\label{fig:MotivatingExample}
\end{figure*}

% \subsection{Static Scheduling} 
\textit{Static HLS} uses modulo scheduling to map operations to cycles at compile time \cite{modulo_sched, modulo_sched_canis, modulo_scheduling_koch}.
To calculate the II of a loop, modulo scheduling goes over all recurrences (inter-iteration dependencies) in its data dependence graph (DDG) and calculates their $delay$  (the number of cycles needed to traverse the whole recurrence path), and its dependence $distance$ (the number of iterations between the definition of a recurrence value and its use).
The final recurrence constrained II is the maximum over all recurrences in the DDG: 
$$II = max_i  \lceil delay_i/distance_i \rceil.$$
Crucially, static scheduling has to arrive at $one$ II for a loop that needs to accommodate all control-flow paths through the DDG.
For example, in the example from fig. \ref{fig:MotivatingExampleCode} there is a recurrence for \texttt{x}.
Even if the \texttt{x > 100} condition would be satisfied only half of the iterations, modulo scheduling needs to allocate cycles for the operations in the \texttt{if} body and will produce the schedule in fig. \ref{fig:MotivatingExampleStaticSchedule}.
In practice,  control-dependent operations are if-converted -- they execute at runtime but their result might be discarded depending on control flow.
% A static schedule has to assume that this dependency occurs on every iteration and thus it cannot start the next iteration of the loop until the previous iteration has committed its store.

% \input{Figures/StaticDynamaticAndThisApproach}

% \subsection{Dynamic Scheduling} 
\textit{Dynamic HLS} uses dataflow scheduling to trigger the execution of operations based on the availability of data, similar to the principles of first dataflow computers \cite{Arvind_Culler_1986}.
This allows the II of a loop to naturally adapt to runtime conditions.
For the example code in fig. \ref{fig:MotivatingExampleCode}, dynamic HLS would produce the ideal schedule from fig. \ref{fig:MotivatingExampleIdealSchedule}. 
However, it would do so at the expense of dynamically scheduling the whole circuit, even if only one part of it exhibits dynamic behavior.
When mapped to FPGA technology, such dataflow circuits often use several times more resources and have a significant critical path overhead compared to static HLS \cite{dynamatic_ltl}.
There is a need to systematically and intelligently combine static and dynamic HLS scheduling.
% Previous work has shown that marking parts of dataflow circuits for static scheduling can be beneficial.
% We show that introducing selective dynamic scheduling into static HLS is also possible.
% To the best of our knowledge, there does not exist an automated compiler analysis which extracts regions of code amenable for dynamic scheduling.
% In this paper, we study the problem of selectively introducing dynamic scheduling into static HLS. 
To this end, we make the following contributions:
\begin{itemize}
    \item A compiler analysis informed by modulo scheduling for discovering basic blocks, memory operations, and whole loops suitable for dynamic scheduling (sec. \ref{sec:analysis}).
    \item A method for the automatic introduction of dynamic scheduling inside modulo scheduled HLS. We show how basic blocks and loops can be transformed into predicated processing elements, and how the decoupled memory access/execute technique can be combined with a Load-Store Queue to achieve out-of-order dynamically scheduled memory operations in static HLS (sec. \ref{sec:transformation}).
    \item An evaluation of our work against three other approaches to HLS scheduling: static, fully dynamic (Dynamatic \cite{josipovic_dynamatic_2022}), and DASS \cite{dass} which introduces static islands into otherwise dynamically scheduled HLS. On a set of ten benchmarks, we show an up to 3.7$\times$ and 3$\times$ speedup on average against Dynamatic and DASS, respectively, while achieving a lower area overhead and critical path overhead (sec. \ref{sec:evaluation}).
\end{itemize}

\section{Background \& Related Work} \label{sec:background}

\subsection{Dynamic Scheduling} \label{sec:backgroundDynSched}

% CASH \cite{Venkataramani2001} was one of the first C to hardware compilers that used dynamic scheduling.
% It differs from recent dynamic HLS in that it used asynchronous hardware.
% CASH used an explicit dataflow Intermediate Representation (IR) called Pegasus IR \cite{Budiu2003}, a form of predicated Single Static Assignment (SSA) \cite{Ottenstein1990ThePD}, which implicitly transforms control-flow into dataflow.

Carloni \textit{et al.} formalized a theory of latency-insensitive design \cite{Carloni_McMillan_Sangiovanni_Vincentelli_2001}.
Assuming that modules are stallable, their protocol separates module communication from their cycle behavior.
Later, Cortadella \textit{et al.} proposed a simplified latency-insensitive protocol called SELF, which could be applied in synchronous circuits \cite{Cortadella_Kishinevsky_Grundmann}.
The SELF protocol, or modifications thereof, has since been used as a basis for dynamically scheduled circuits \cite{Kam_Kishinevsky_Cortadella_micro_pipelining, huang2013elastic, elastic_flow,josipovic_dynamatic_2022, Townsend_Kim_Edwards_2017}.
Most commercial HLS tools provide a latency-insensitive channel construct, e.g. SYCL Pipes \cite{sycl_khronos}, or Xilinx Streams \cite{vitis}.
We use SYCL pipes in this work to introduce dataflow scheduling across kernels.

Several HLS tools that automatically create dataflow circuits have been developed in academia \cite{josipovic_dynamatic_2022,elastic_flow,Edwards2019CompositionalDC,Venkataramani2001}.
Dynamatic by Josipović \textit{et al.} \cite{josipovic_dynamatic_2022} is the most recent and most general of them, introducing latency-insensitivity for every def-use SSA value pair that spans across two basic blocks.
The resulting circuits perform well on irregular code, but they use more dynamic scheduling than required.
Instead, we selectively introduce the minimum amount of dynamic scheduling to achieve the same throughput by using the Data Dependence Graph (DDG) and Control Dependence Graph (CDG) to connect selected dynamically scheduled producer and consumer pairs directly.
In contrast, Dynamatic materializes dataflow constructs using the Control Flow Graph (CFG) \cite{Rosen1988GlobalVN} order, which means that a token needs to flow through all basic blocks between a producer and consumer (although recent work started to address this issue \cite{dynamatic_gated_ssa}).

The work of Cheng \textit{et al.} shares our goal of combining dynamic and static scheduling \cite{dass, cheng_finding_static}.
They extended Dynamatic with the DASS methodology (Dynamic and Static Scheduling) \cite{dass}, which lets programmers manually identify static islands in their otherwise dynamically scheduled circuit.
Later, the authors provided formal guidelines on where static islands can be beneficial \cite{cheng_finding_static}.
Static islands can be scheduled statically on the inside and are wrapped in interfacing logic to communicate results with the dynamic part of the circuit.
The performance and resource usage of such a hybrid approach is promising, but the circuit critical path is still bottlenecked by the dynamic part.
Furthermore, some of the restrictions on what can be marked as a static island are prohibitive, e.g. Load Store Queue (LSQ) connections can only be made from the dynamic part and an additional memory controller is needed to arbitrate between the static and dynamic region if they access the same part of memory \cite{cheng_finding_static}.
Our approach has no such restriction.
A major difference between DASS and our work is that we propose to introduce dynamic scheduling into modulo-scheduled HLS using only constructs available in static HLS.

Xu \textit{et al.} \cite{dynamatic_ltl} proposed to use linear temporal logic (LTL) to prove that certain handshaking signals in a dataflow circuit will never be used or that they are equivalent to other signals at the time of use, allowing them to be removed without losing correctness.
While this brings the resource usage of dataflow circuits closer to static HLS, there is still a significant gap between the two, both in terms of resource usage and achievable critical path.
Furthermore, model checking of LTL formulas is notorious for its exponential complexity in the number of transitions in the system, e.g. \cite{dynamatic_ltl} reports 80 min check time for a code with two matrix multiply loops.
The authors use the abstraction technique to reduce the size of the state system, but it remains to be seen how this approach performs on more complex codes that result in a DDG with high connectivity (and thus more states that cannot be abstracted away).
We propose to tackle the problem of resource usage by selectively introducing dynamic scheduling into static HLS.
% We argue that the number of dynamic regions in codes is much smaller than the number of static regions, making it more beneficial to selectively introduce dynamic behavior into static HLS than vice versa.
% We also argue that it's easier to find dynamic regions starting from modulo-scheduled HLS than to recover static information in a fully latency-insensitive system.

% There is also a line of work that aims to improve the quality of results from static scheduling by using source-to-source transformations \cite{speculative_source_to_source, liu_pipelining_polly}.
% For example, Liu \textit{et al.} \cite{liu_pipelining_polly} applied polyhedral analysis to split the iteration space based on the minimum II that they can safely use, generating a separate loop for each split.
% % Derrien \textit{et al.} \cite{speculative_source_to_source} proposed to improve loop pipelining on unpredictable codes by speculative execution realized via a source-to-source transformation.
% Source-level approaches are usually tightly coupled to the specific HLS tool that is being targeted.
% Our analysis and transformations are expressed at the Control and Data Dependency Graph level and informed by the fundamental limitation of modulo scheduling, making them independent of any particular HLS tool.

Our approach to dynamic scheduling resembles decoupled software pipelining (DSWP) proposed by Ottoni \textit{et al.} \cite{dswp}.
In DSWP, the DDG and CDG of a loop are partitioned into strongly connected components (SCCs), which are decoupled into separate threads communicating via FIFOs.
Although originally proposed for multicore CPUs, the decoupling approach works especially well on FPGAs where FIFO communication is efficient, e.g. S. Cheng \textit{et al.} used the DSWP principle to minimize stalls resulting from cache misses on reconfigurable accelerators \cite{decoupled_memory_wawrzynek}.
% Chen \textit{et al.} presented an architectural template for memory prefetching by decoupling memory accesses from the compute pipeline \cite{decoupled_memory_prefetching}.
Although similar in nature, our approach is fundamentally different because our goal is the selective introduction of dynamic scheduling and we perform the decoupling \textit{inside an SCC}, while DSWP decouples \textit{whole SCCs}. 
To illustrate the difference, consider the DDG and CDG in fig. \ref{fig:MotivatingExampleDDG}.
DSWP would decouple the whole recurrence SCC $1 \rightarrow 2 \rightarrow 3$, while we would decouple only node $2$.

\subsection{Dynamic Dataflow Model of Computation in Static HLS}

% The original OpenCL 1.2 specification (cite) was not a good match for the spatial parallelism of FPGAs -- its focus was on data parallelism for GPUs.
% The newer 
% In SDF, the required FIFO sizes between kernels can be inferred at compile time because the consumption/production rates of kernels are known.
% Pipe reads/writes can be specified as blocking or non-blocking.
% Consequently, they are well suited for the acceleration of regular dataflow codes (e.g. stencils \cite{stencil-flow}) where 

% In SDF, the programmer creates processing elements (OpenCL kernels) and connects them using FIFOs (OpenCL pipes).
% Crucially, the consumption and production rates of every processing element must be computable at compile-time.
% This restriction allows for statically determining the compute schedules and buffer sizes for the FIFO pipes.
% The SDF model is a good fit for regular codes, like neural networks or stencil computations, where enough compile-time information is provided to create a static schedule \cite{best_HLS_practices}.
% It is not suited irregular codes, where the production/consumption rates depend on runtime information.
% (concepts like dynamic connectivity of pipes are not supported).
% Both AMD and Intel have SYCL implementations called TriSYCL and DPC\texttt{++}, respectively.
% We use DPC\texttt{++} in this paper, but our work applies equally well to AMD TriSYCL.
% In fact, TriSYCL is built on top of DPC\texttt{++}.
% \cite{Lee1987SynchronousDF} 

% \input{Figures/PipeDiagram}

We use SYCL \cite{sycl_khronos} HLS as a representative of modulo-scheduled HLS in this work.
Each SYCL kernel has its own static schedule, and kernels can communicate with each other via latency-insensitive SYCL pipes.
In previous work, pipes were used to implement the Synchronous Dataflow (SDF) model of computation \cite{opencl_moc}, which guarantees deterministic execution by enforcing compile-time computability of pipe read/write rates.
To enable dynamic scheduling the Dynamic Dataflow (DDF) model of computation is needed -- pipe read/writes should be allowed to depend on the program control-flow \cite{Hauck_DeHon_2008}. 
This is possible in SYCL HLS, allowing the construction of the switch/select dynamic dataflow primitives.
In the context of our work, SYCL is advantageous over OpenCL because SYCL pipes are implemented as types, not kernel arguments, making them easier to use in compiler transformations.
Furthermore, there is ongoing work on an MLIR \cite{mlir} SYCL dialect to make kernel fusion and fission a first-class compiler transformation \cite{sycl_kernel_fusion}.
To the best of our knowledge, there is no prior work that shows that the DDF model of computation can be achieved in static HLS.

SYCL pipe operations can be blocking or non-blocking.
When using a non-blocking pipe operation, the pipe returns a \texttt{success} code depending on whether the operation was completed, without stalling the pipeline.
A kernel can then make control flow decisions based on the availability of data in the pipe.
This behavior enables the Dynamic Dataflow (DDF) with Peeks model of computation \cite{Hauck_DeHon_2008}.
Compared to vanilla DDF, DDF with Peeks is non-deterministic.
Non-determinism can be used to, for example, implement a Load-Store Queue (LSQ) connected to a variable latency memory system.
% However, spec LSQs are beyond the scope of this paper and we make sure to leave any LSQ effects out of the evaluation.

\section{Compiler Discovered Dynamic Scheduling}  \label{sec:analysis}

This section presents how to find basic blocks, memory operations, or whole loops amenable to dynamic scheduling.

\subsection{Marking Basic Blocks}

A loop schedule in static HLS is obtained using modulo scheduling \cite{modulo_sched, modulo_sched_canis, modulo_scheduling_koch}, which arrives at a minimum recurrence-constrained loop initiation interval (II) by taking the maximum over all SCCs in the DDG:
$$II = max_i  \lceil delay_i/distance_i \rceil,$$
where the $delay$ is the sum of instruction latencies on the SCC path, and $distance$ is the minimum iteration distance between the definition of the value calculated by the SCC and its use.
% Note that the nodes in a DDG are individual instructions, not basic blocks of instructions.
Since modulo scheduling has to arrive at a single II, it has to necessarily over-approximate the recurrence-constrained II if there are control-dependent paths through the DDG with a lower $delay$ or a higher $distance$.
The key idea of this paper is to selectively decouple parts of the SCC into separate modulo-scheduling problems, such that modulo scheduling doesn't have to over-approximate. 

Alg. \ref{alg:markingBBs} describes our analysis for marking basic blocks for dynamic scheduling.
We propose to enumerate all possible control-flow paths through the SCC and calculate their II.
For each SCC path with an $II > 1$, we collect instructions that are control dependent on anything else but the loop header (every instruction inside a loop body is trivially control-dependent on the loop header).
For every collected DDG node, we obtain all other DDG nodes from the same basic block and calculate their contribution to the II of the currently considered path.
Specifically, we check if without the collected nodes the path $delay$ decreases or the dependence $distance$ increases.
If true, we mark that block for dynamic scheduling and collect all instructions in the block that are part of the currently considered SCC path.
The block could contain instructions that are not part of the currently evaluated SCC in which case they will not be marked, or they will be marked when evaluating a different SCC.
One could set a threshold for the $delay$ decrease or dependence $distance$ increase (e.g. to avoid dynamic scheduling overhead if the II improvement is small), however, this is beyond the scope of this paper.
% Instructions from the same block marked for dynamic and belonging to different SCCs will be clustered together.

\algrenewcommand\algorithmicindent{1em}%

\begin{algorithm}[t]
\caption{Marking Basic Blocks for Dynamic Scheduling}\label{alg:markingBBs}
\begin{algorithmic}
\State \textbf{Input: } DDG, CDG, CFG
% \State \textbf{Output: } List of Basic Blocks \textit{BBs}
\For{$SCC \in DDG$}
    \For{every legal control-flow $Path \in SCC$}
    \State $PathII \gets CalculateII(Path)$
    \State \textbf{if} $PathII = 1$ \textbf{then} continue
    % \If{$PathII > 1$}
        \For{$Node \in Path$}
            \State $BB \gets BasicBlock(Node)$
            \State $NodesBB \gets AllDDGNodes(BB)$
            \State $C_1 \gets CtrlDepSrc(BB) \neq LoopHeader$
            \State $C_2 \gets CalculateII(Path \setminus NodesBB) < PathII$
            \If{$C_1$ \textbf{and} $C_2$}
                % \State $BBs \gets BBs \cup BB$
                \State \textbf{mark} $BB$ \textbf{for dynamic scheduling}
            \EndIf
        \EndFor
    % \EndIf
    \EndFor
\EndFor
\end{algorithmic}
\end{algorithm}

            % \State $NodesBB \gets \{n \mid n \in DDG and getNodeBB(n) == BB\}$

\textit{Example:} Consider the DDG in fig. \ref{fig:MotivatingExampleDDG} with two SCCs: $(1 \rightarrow 4)$ and $(1 \rightarrow 2 \rightarrow 3)$.
$(1 \rightarrow 4)$ has a trivial II of 1, so it is not marked.
The second SCC has an II of 5, so we check if it contains any control-dependent nodes by using the CDG \cite{Ferrante1987ThePD}.
The blocks containing nodes $1, 3$ are control-dependent on the loop header block, so they are ignored.
The block containing node $2$, however, is control-dependent on a non-loop-header block, so it is marked for dynamic scheduling.

\subsection{Marking Memory Operations}

Memory operations that cannot be disambiguated at compile time form memory-dependency edges in the DDG \cite{Ferrante1987ThePD}.
Modulo scheduling treats these edges in the same way as it treats register dependencies.
The only difference is that the dependence $distance$ between memory operations can be unknown, for example, if the access pattern is data-dependent or the compiler doesn't employ a strong enough alias analysis \cite{dependence_distance_hls, poly_hls}.
If the dependence distance is known, we employ the same strategy as for marking basic blocks, namely, we check if there is a control flow path through the DDG with a higher dependence $distance$, and if yes, we check if it's control dependent on anything but a loop header.
If the dependence distance between dependent memory operations is unknown, we immediately mark them for dynamic scheduling.
For any marked pair of memory operations, we also mark all other memory operations that use the same base pointer.

\subsection{Inter-Loop Pipelining}

% So far, we have collected instructions for dynamic scheduling for cases where modulo scheduling fundamentally has to arrive at a worst-case schedule.
% In addition to dynamically scheduling instructions inside a single loop, our methodology extends to whole loops.
The opportunities for decoupling whole loops are rare.
The first scenario is a nested control-dependent loop that is control-dependent on anything else than its parent loop header, and where the parent loop is not perfectly pipelined because of a dependency in the nested loop.
That is, the decoupling of the inner loop should improve the average II of the outer loop.
% This scenario is rare, because in most HLS applications the performance depends on the pipelining of inner, not outer, loops.

The second opportunity for decoupling whole loops is a scenario with multiple sibling loops -- loops at the same level of nesting.
If a loop $L_1$ has a sibling loop $L_2$, then we check if it's legal to start the second loop before the first one has finished.
We mark $L_2$ for dynamic scheduling if:
\begin{enumerate}
    \item There are no data dependencies between $L_1$ and $L_2$ calculated by a recurrence, and with a source in $L_1$ and destination in $L_2$. In other words, if the dependency destination in $L_2$ needs to wait for the whole $L_1$ to finish, then there is no benefit to decoupling $L_2$.
    \item There are no memory dependencies between $L_1$ and $L_2$ such that the address expressions in $L_1$ and $L_2$ cannot be disambiguated at compile time.
\end{enumerate}

Compile time memory disambiguation across loops is often not possible.
For example, in the polyhedral model \cite{poly_cc} it would require proving that the $L_1$ and $L_2$ polyhedra do not overlap at all.
Connecting memory operations in $L_1$ and $L_2$ loops to an LSQ would be of little benefit because the $L_2$ loop would have to wait for all allocations in $L_1$ to finish.
This has also been noted by Cheng \textit{et al.} \cite{dynamic_inter_block_cheng} who proposed to statically disambiguate memory accesses across loops for individual iterations, rather than the whole iteration space.
The idea is that individual iterations of the second loop can start as soon as possible, while iterations with offending memory operations will stall.
This is a promising approach and could be integrated into our flow.
However, codes amenable to dynamic scheduling often have data-dependent address expressions, making the applicability of this approach limited.
Future work could investigate if the approach can be extended to a lightweight runtime mechanism, similar to the work of Liu \textit{et al.} \cite{online_poy_hls}.
% to special cases.
% for inter-loop memory disambiguation of individual loop iterations.

% .
% If both loops write to the same memory addresses and the expressions calculating the memory addresses in $L_1, L_2$ cannot be proven not to alias at compile time, then we mark those memory operations for dynamic scheduling, i.e. an they will be connected to an LSQ.
% Compile time memory disambiguation across loops is often not possible, even using the polyhydral model \cite{poly_cc}, since it requires proving that the $L_1$ and $L_2$ polyhedra do not overlap at all.
% On the other hand, if the memory operations don't need an LSQ in loops $L_1$ and $L_2$ on their own, then inserting an LSQ is often a large expense for little benefit. 
% This is because, in most cases, the $L_2$ would have to wait for all memory operations in $L_1$ to finish.
% Cheng \textit{et al.} \cite{dynamic_inter_block_cheng} studied the problem of statically disambiguating individual iterations of the second loop.
% However, the approach is limited if the address expressions are data-dependent, which is often the case in codes amenable to dynamic scheduling.
% Future work could investigate lightweight runtime mechanisms for inter-loop memory disambiguation that do not require the large overhead of an LSQ.
% % All is needed in this case is signalling individual iterations of the second loop that it is safe to execute.

\section{Achieving Selective Dynamic Scheduling} \label{sec:transformation}

This section presents our main contribution: a method for introducing dynamically scheduled code regions in modulo-scheduled HLS via latency-insensitive channels.

\subsection{Dynamically Scheduled Basic Blocks} \label{sec:dynSchedBB}

\begin{figure}[t]
\centering
\subfloat[Statically scheduled loop with a worst case II.]{\includegraphics[width=0.20\textwidth]{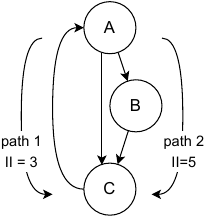}\label{fig:introCFGa}}%
\hfill%
\subfloat[Decoupled unpredictable control-dependent data dependency.]{\includegraphics[width=0.235\textwidth]{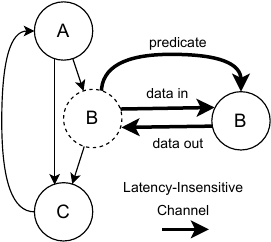}\label{fig:introCFGb}}%
% \vfill%
% \subfloat[Our work: decoupled unpredictable memory accesses.]{\includegraphics[width=0.45\textwidth]{Figures/Intro_cfg_c.pdf}\label{fig:introCFGc}}%

\caption{Our main idea. Control-flow paths with a higher recurrence-constrained initiation interval (II) are decoupled into separate modulo scheduled instances, with data dependencies communicated via latency-insensitive channels. A recurrence through registers is decoupled into a predicated PE.}
% ; a recurrence through memory is decoupled into a Load Store Queue and address generation units (c).

\label{fig:introCFG}
\end{figure}

Basic blocks marked for dynamic scheduling are transformed into predicated Processing Elements (PEs).
Fig. \ref{fig:introCFGa} shows a possible CFG for our motivating example code from fig. \ref{fig:MotivatingExample}.
Fig. \ref{fig:introCFGb} shows a high-level overview of how the marked block \textit{B} would be decoupled by our transformation.
All instructions collected in the marked block are moved from the original CFG to the predicated PE.
We then collect the set of input and output data dependencies between the PE and the original CFG using a simple data flow algorithm: every SSA value used in the PE but defined in the original CFG is an input dependency from the original CFG to the PE, and vice versa for output dependency from the PE to the original CFG.
% Additionally, if the block terminator branch uses an SSA value calculated in the predicated PE, then that value is also collected as an output dependency.
All SSA values collected as input dependencies are replaced with pipe writes in the original CFG, and with pipe reads in the PE.
The dual is done for output dependencies.
Finally, we insert a predicate pipe write to the beginning of the decoupled block in the original CFG which will trigger our predicated PE whenever control transfers to that block.

The PE is guaranteed not to access any memory directly. 
If a memory access inside a marked block was itself marked for dynamic scheduling, then it will be replaced by pipe read or write (subsec. \ref{sec:dynSchedMem}).
If the access was not marked, we keep it in the original CFG and communicate its operands as dependencies between the PE and the original CFG -- a load used in the PE becomes an input dependency, a store operand defined in the PE becomes an output dependency.
% its operand or result will become an output or input dependency, respectively, into the predicated PE.

As presented so far, our transformation is local to a basic block and doesn't require updating SSA values in other blocks.
This can change after \textit{hoisting redundant pipe operations} out of loops.
A pipe operations in the main CFG can be hoisted out before or after the loop if the value it is carrying is only used or defined in the predicated PE.
For example, the code in fig \ref{fig:MotivatingExampleCode} would not have any pipe operations hoisted out, because the \texttt{x} value would be used in both the PE and the original CFG.
If, however, the \texttt{x += g(x)} statement would be removed, then the pipe operations supplying and receiving \texttt{x} could be hoisted out because the original CFG would not use its value in the communication sequence: $CFG \xrightarrow{x} PE \xrightarrow{x} CFG \xrightarrow{x} PE$.

\textit{Effect of transformation:} Since pipe operations do not form inter-iteration dependencies, modulo scheduling will find that the fig. \ref{fig:MotivatingExample} loop $delay$ is now 3 and not 5.
Whenever control transfers to the decoupled basic block, the original loop will trigger the predicated PE and communicate the required input dependencies.
It will then continue its execution until it encounters an operation that is dependent on an output dependency from the decoupled block.
If such a dependency is encountered, then the original loop is stalled until the required dependency is communicated from the PE.
Thus, a variable II is achieved which naturally adapts to runtime conditions.

\textit{Dynamic scheduling of whole loops} is achieved in the same fashion as for basic blocks, with the difference that the dependencies are calculated for the whole loop.

\subsection{Dynamically Scheduled Memory Operations} \label{sec:dynSchedMem}

\begin{figure}[t]
\centering
\includegraphics[width=0.43\textwidth]{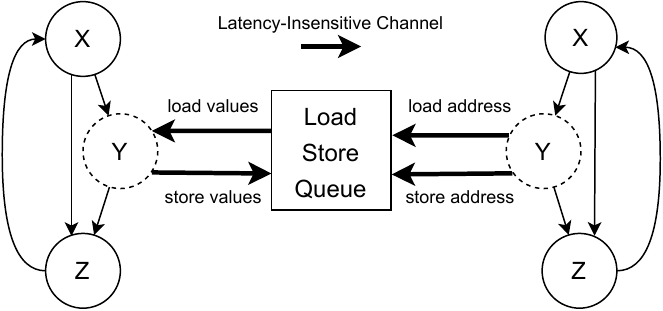}\label{fig:introCFGc}

% \caption{A recurrence through memory is decoupled into a Load Store Queue and address generation units.}
\caption{An illustration of how a recurrence through memory is decoupled.}

\label{fig:introCFG_LSQ}
\end{figure}

% Dynamically scheduling memory operations in static HLS is more difficult than dynamically scheduling basic blocks of instructions.
Memory operations marked for dynamic scheduling require runtime memory disambiguation machinery, such as a Load-Store Queue (LSQ) \cite{Josipovic_Brisk_Ienne_2017}.
An LSQ can check for memory conflicts at runtime by comparing load and store addresses out-of-order with the actual memory accesses, and stall the datapath if a true data hazard is detected.
One can easily plug any LSQ design into modulo-scheduled HLS, however, this is not enough.
For an LSQ to be most effective, it should be able to accept load and store requests (address allocations) ahead of store values.
In a dataflow circuit, this happens naturally since the production of memory addresses is decoupled from the actual load and store operations.
To achieve the same effect in modulo-scheduled HLS, the address generation should also be decoupled, similar to the principle of decoupled access/execute architectures \cite{decoupled_access_exec}.
Decoupled memory accesses have been studied before in the context of FPGAs, but only for prefetching and hiding variable latency memory accesses \cite{decoupled_memory_prefetching, decoupled_memory_wawrzynek}.
We contribute the insight that this approach, together with an LSQ, can enable dynamically scheduled out-of-order loads in static HLS.
Before describing the actual transformation, we give an algorithm for automatically checking if decoupled address generation can run ahead of memory accesses.

Given a set of address-generating instructions $I_{GEN}$ for a given base address and a set of memory access instructions $I_{ACCESS}$ using addresses generated by $I_{GEN}$, we decide to decouple the $I_{GEN}$ instructions  if:
\begin{itemize}
    \item[] $\forall i \in I_{GEN}$ where $i$ is used by a store, $i$ is not control nor data dependent on any instruction $j$, such that there is a DDG path from an instruction $k \in I_{ACCESS}$ to $j$.
\end{itemize}
In other words, if the execution of a store, or its address calculation, depends on the value of a load from the same base address, then decoupling is not possible.
An example would be bubble sort, where the store is conditional on the loaded values.
In such codes, LSQ store requests for a given iteration can only be issued once the loads from the previous iteration have finished.
This restriction is not a limitation of our approach, since even a fully dynamic HLS tool would have to stall in such a situation.
Value-based disambiguation, as opposed to an LSQ, might perform better on these codes, because both the load and store could be executed speculatively \cite{Thielmann_Load_Speculation}.

If our analysis determines that decoupling of address generation is profitable, then we proceed with decoupling of the memory-generating instructions.
We copy the original loop CFG and delete from it all instructions not needed by $I_{GEN}$ (these can be easily obtained by walking the DDG).
Pipes for input and output dependencies are materialized similarly to section \ref{sec:dynSchedBB}.
Regardless of whether the generation of memory addresses is decoupled or not, we insert the required pipe calls to supply load and store requests to the LSQ and to supply to it and receive from it store and load values.
Fig. \ref{fig:introCFG_LSQ} shows the resulting communication pattern if the address generation is decoupled.
Load and store instructions in block \textit{Y} have been replaced with latency-insensitive channel reads and writes from and to an LSQ, respectively.
The addresses to the LSQ are supplied by a separate modulo-scheduled component, which contains only address-generating instructions.
The generation of load and store addresses in this decoupled component is control-flow equivalent to the consumption and generation of load and store values in the original CFG.

\subsection{Composability of Transformations} \label{sec:composability}

The presented transformations are composable. % for introducing dynamically scheduled basic blocks, loops, and memory operations
A decoupled loop can have a number of its own basic blocks decoupled, and the basic blocks can have dynamically scheduled memory operations.
The problem of LSQ request ordering across decoupled code regions is solved by the design of our LSQ.
Our LSQ is based on tagged memory operations -- each load and store request is tagged with an integer value which represents the state of the memory at that point; stores increment the tag before making a request, loads use the tag directly.
The function of the tag inside the LSQ is beyond the scope of this paper, but it in effect produces a data dependency chain between memory stores and other memory operations (similar to what Elakhras \textit{et al.} proposed for the Dynamatic LSQ \cite{straight_to_the_queue}).
This tag dependency chain is picked up through our input and output dependency collection, and as a result is communicated between decoupled code regions according to runtime control flow, naturally taking care of the correct order of LSQ requests.
% Fig. \ref{fig:FullTransformation} shows how our transformations compose on a real code (\textit{getTanh}).

% \input{Figures/FullMuxTransform}

\begin{figure}[t]
\centering
\includegraphics[width=0.49\textwidth]{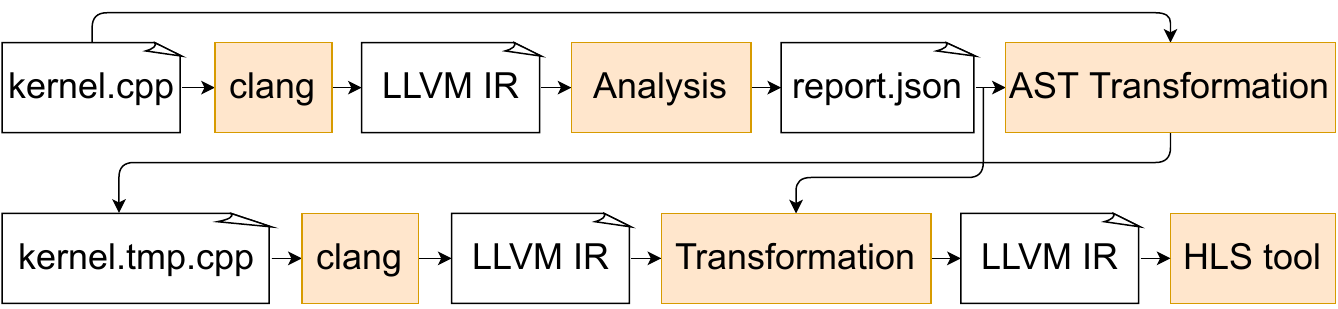}

\caption{Our tool flow. The analysis and transformation parts are described in the paper (sec. \ref{sec:analysis} and \ref{sec:transformation}). The AST transformation consists of creating kernel copies, inserting LSQ kernels and creating latency-insensitive channels, which are later used by the transformation operating on the LLVM IR.}

\label{fig:ToolFlowSimple}
\end{figure}

\section{Evaluation} \label{sec:evaluation}

\definecolor{seabornBlue}{RGB}{76,114,176}
\definecolor{seabornGreen}{RGB}{85,168,104}
\definecolor{seabornRed}{RGB}{196,78,82}

\begin{figure*}[t!]

% \subfloat[]{%
\hspace{8pt}%
\begin{tikzpicture}
\begin{axis}[
    x tick label style={font=\footnotesize},
    y tick label style={font=\footnotesize},
    width=\textwidth,
    height=4cm,
	% x tick label style={
	% 	/pgf/text sep=},
    symbolic x coords={sparseMatrix, getTanhDouble, filterSum, vecNormTrans, histogram, bubbleSort, getTanh, BNN, covariance, gesummv, gesummv2},
	ylabel={Area $\times$SS},
    ymin=0, ymax=8,
    ytick={1,2,3,4,5,6,7,8},
	enlargelimits=0.01,
	legend style={at={(0.5,1.3)},
	anchor=north,legend columns=-1},
	ybar interval=0.5,
    ymajorgrids=true,
]
\addplot[fill=seabornRed, color=seabornRed] coordinates {(sparseMatrix,2.5) (getTanhDouble,1.1) (filterSum,2.8) (vecNormTrans,2.6) (histogram, 2.1) (bubbleSort, 7.1) (getTanh,3.5) (BNN,6.8) (covariance,3.4) (gesummv,1.6) (gesummv2,1.6) };

\addplot[fill=seabornGreen, color=seabornGreen] coordinates {(sparseMatrix,1.3) (getTanhDouble,1.2) (filterSum,1.7) (vecNormTrans,2.4) (histogram, 2.4) (bubbleSort, 7.5) (getTanh,1.3) (BNN,2.9) (covariance,1.6) (gesummv,1.3) (gesummv2,1.6) };

\addplot[fill=seabornBlue, color=seabornBlue] coordinates {(sparseMatrix,1) (getTanhDouble,1.5) (filterSum,2) (vecNormTrans,2) (histogram, 1.2) (bubbleSort, 1.6) (getTanh,1.7) (BNN,1.4) (covariance,1) (gesummv,1) (gesummv2,1.6) };

\legend{Dynamatic,DASS,This Work}
\end{axis}
\end{tikzpicture}

% \subfloat[]{%
\vspace{-0.4cm}
\begin{tikzpicture}
\begin{axis}[
    x tick label style={font=\footnotesize},
    y tick label style={font=\footnotesize},
    width=\textwidth,
    height=4cm,
    symbolic x coords={sparseMatrix, getTanhDouble, filterSum, vecNormTrans, histogram, bubbleSort, getTanh, BNN, covariance, gesummv, gesummv2},
	ylabel={Speedup $\times$SS (log scale)},
    ymode=log,
    ytick={0, 0.1, 0.2, 0.5, 1, 2, 5, 10, 20, 40},
    log ticks with fixed point,
	enlargelimits=0.01,
	% anchor=north,legend columns=-1},
    ymajorgrids=true,
	ybar interval=0.5
]

\addplot[fill=seabornRed, color=seabornRed, opacity=1, error bars/.cd, y dir=minus, y explicit, error bar style={opacity=1, color=seabornRed, mark size=3pt, line width=2pt, xshift=7.5pt}] coordinates {(sparseMatrix,2.26) -= (0,1.88) += (0,0) (getTanhDouble,16.52) -= (0,0) += (0,0) (filterSum,2.19) -= (0,1.75) += (0,0) (vecNormTrans,0.97) -= (0,0) += (0,0) (histogram, 3.72) -= (0,0) += (0,0) (bubbleSort, 0.62) -= (0,0.31) += (0,0) (getTanh,5.71) -= (0,5.44) += (0,0) (BNN,0.55) -= (0,0) += (0,0) (covariance,0.18) -= (0,0) += (0,0) (gesummv,0.07) -= (0,0) += (0,0) (gesummv2,1)  -= (0,0) += (0,0)};

\addplot[fill=seabornGreen, color=seabornGreen, opacity=1, error bars/.cd, y dir=minus, y explicit, error bar style={opacity=1, color=seabornGreen, mark size=3pt, line width=2pt, xshift=23pt}] coordinates {(sparseMatrix,2.26) -= (0,1.87) += (0,0) (getTanhDouble,16.52) -= (0,0) += (0,0) (filterSum,2.23) -= (0,1.78) += (0,0) (vecNormTrans,0.96) -= (0,0) += (0,0) (histogram, 3.72) -= (0,0) += (0,0) (bubbleSort, 0.61) -= (0,0.31) += (0,0) (getTanh,5.71) -= (0,5.39) += (0,0) (BNN,0.53) -= (0,0) += (0,0) (covariance,0.19) -= (0,0) += (0,0) (gesummv,0.24) -= (0,0) += (0,0) (gesummv2,1)  -= (0,0) += (0,0)};

\addplot[fill=seabornBlue, color=seabornBlue, opacity=1, error bars/.cd, y dir=minus, y explicit, error bar style={opacity=1, color=seabornBlue, mark size=3pt, line width=2pt, xshift=38.5pt}] coordinates {(sparseMatrix,14.33) -= (0,13.52) += (0, 0) (getTanhDouble,37.93)  -= (0,0) += (0,0)  (filterSum,4.92) -= (0,3.95) += (0,0) (vecNormTrans,1.99)  -= (0,0) += (0,0)  (histogram, 4.22)  -= (0,0) += (0,0)  (bubbleSort, 0.75)  -= (0,0.37) += (0,0)  (getTanh,19.35)  -= (0,18.89) += (0,0)  (BNN,2.33)  -= (0,0) += (0,0)  (covariance,1)  -= (0,0) += (0,0)  (gesummv,1)  -= (0,0) += (0,0) (gesummv2,1)  -= (0,0) += (0,0) };

\legend{}
\end{axis}
\end{tikzpicture}

\caption{Area overhead and speedup of Dynamatic \cite{josipovic_dynamatic_2022}, DASS \cite{dass} and this work against their respective statically scheduled (SS) baselines. The range bars in the speedup plot represent the range of speedup as the data distribution changes. A speedup below 1 indicates a slowdown relative to static scheduling.}

\label{fig:AreaAndPerfPlot}

\end{figure*}
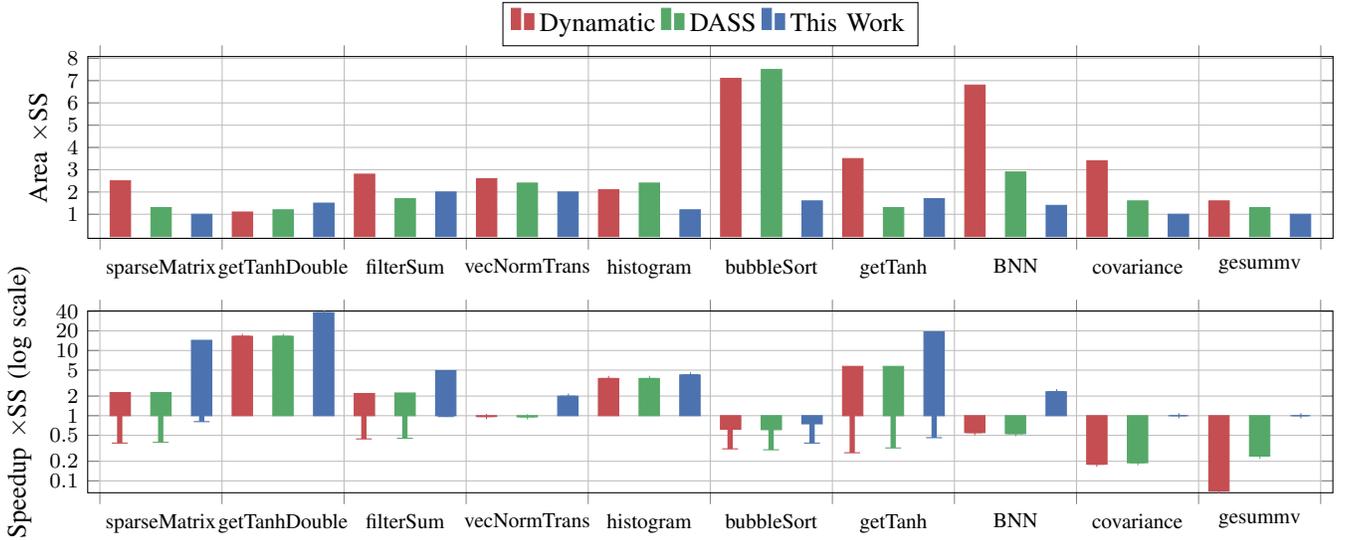

\subsection{Methodology}

We implemented our compiler analysis and transformations in the LLVM framework \cite{LLVM_CGO04} and integrated them with the Intel SYCL compiler.
Fig. \ref{fig:ToolFlowSimple} shows an overview of our tool flow.
Our implementation is publicly available.\footnote{\url{https://github.com/robertszafa/elastic-sycl-hls}}
We evaluate our work against three other scheduling approaches: static (SS), fully dynamic (DS) \cite{josipovic_dynamatic_2022}, and DASS \cite{dass}, which allows the marking of a function for static scheduling inside Dynamatic.
Dynamatic is based on Xilinx tools, while our approach is implemented on top of Intel HLS, which makes a direct comparison in terms of absolute area usage difficult.
Because of that, we compare the \textit{normalized} area usage of Dynamatic, DASS, and our approach against their \textit{respective} static HLS baseline.
The register and LUT usage overhead are combined using geometric mean into a single area overhead.
For Dynamatic and DASS we used the post-synthesis report from Vivado 2020.2 for the Xilinx xc7k160tfbg484; our approach used Quartus 19.2.0 post-synthesis reports for the Altera 10AX115S. % Kintex 7 part (xc7k160tfbg484), Zynq part (xc7z020clg484
% We did not set any critical path constraints.
Clock cycles were obtained using ModelSim.

We applied our approach to ten benchmarks from the HLS literature \cite{dass, josipovic_dynamatic_2022} made publicly available by Cheng \textit{et al.} \cite{benchmarks_jianyi_cheng_2019}.
Where applicable, we report worst- and best-case performance for different input data distributions.
\textit{sparseMatrixPower} has a control-dependent nested loop.
\textit{getTanhDouble}, \textit{filterSum} and \textit{vecNormTrans} are single loops that have recurrences with control-dependent parts.
\textit{histogram}, \textit{bubbleSort}, \textit{getTanh} and \textit{BNN} have memory accesses with unpredictable addresses.
We also include two codes without any dynamic behavior: a matrix \textit{covariance} calculation and \textit{gesummv} which is a scalar, vector, and matrix multiply.
In codes with unpredictable memory addresses, we use an LSQ adapted to our approach, while Dynamatic and DASS use the Dynamatic LSQ \cite{Josipovic_Brisk_Ienne_2017}.
Any difference in the experiments due to the different LSQ designs is left out of the evaluation --  we don't include the LSQ areas, while the throughput of the two LSQs is the same.
All codes use on-chip memory, although we support off-chip memory addresses equally well.

\subsection{Results}

Fig. \ref{fig:AreaAndPerfPlot} shows the area overhead and speedup over static scheduling of the fully dynamically scheduled approach, DASS, and our work.
Tab. \ref{table:benchmarkTable} features detailed results of all ten benchmarks, which we analyze in the next paragraphs.

\textit{Area:} 
Dynamic scheduling incurs area overhead for handshaking logic, and the missed opportunity for resource sharing: if two hardware components become decoupled via latency-insensitive channels, then the HLS tool cannot make as many latency assumptions as it could if the two components were following the same static schedule.
On average, our approach increases area usage by a factor of $1.3\times$, compared to $2.7\times$ for DS and $1.8\times$ for DASS.
In cases where dynamic scheduling is not beneficial, our approach doesn't make any changes to the static hardware resulting in no resource usage increase, while DS and DASS see a resource increase of $2.3\times$ and $1.4\times$, respectively.
The area overhead of DS and DASS in the \textit{bubbleSort} and \textit{BNN} benchmark is the highest.
\textit{BNN} consists of bit-level logic which we could get more aggressively optimized in the more mature static HLS tools compared to Dynamatic.
\textit{bubbleSort} on the other hand has a large number of basic blocks compared to instructions in them, resulting in a large ratio of dataflow components to functional units in the Dynamatic generated circuit.
We also note that DS and DASS use on average $1.6\times$ and $1.1\times$ more DSPs than static scheduling, while our approach doesn't increase DSP usage.

\textit{Critical path:}
The biggest advantage of our approach is is the higher frequency achievable by static HLS compared to Dynamatic.
On codes 1-4, which don't require an LSQ, our approach results in only an $0.94\times$ frequency drop, compared to $4\times$ frequency reduction for DS and DASS.
On codes 5-8, the critical path is increased significantly by the LSQ in all three approaches.
Future work could investigate alternative LSQ designs with a lower critical path \cite{mem_disamb_approaches}.
On codes without dynamic behavior, DS and DASS see frequency drops, while our approach does not change the SS hardware.

\textit{Throughput:} 
We achieve the same or better throughput as SS, DS, and DASS.
We perform better than DS and DASS on \textit{getTanh} and \textit{BNN} because they involve nested non-trivial control-dependent loops, which benefit from static scheduling.
DASS performs better than Dynamatic on \textit{getTanh} when the data distribution favors static scheduling, but it cannot achieve perfect pipelining when there are no data hazards, because it cannot start the next iteration of the outer loop until the inner loop has returned from its static island.
On codes without any dynamic behavior DS and DASS incur non-trivial overheads, while our approach doesn't change the SS hardware.
% Our communication pattern optimization (sec. \ref{sec:composability}) ensures that the decoupled inner loop directly communicates its result to the LSQ, allowing the outer loop to start sooner.

\textit{Execution time} is the product of the number of cycles and circuit frequency, and since we benefit from the high frequency of SS while achieving the same (or higher) throughput, our approach performs better than DS and DASS.
Across the ten benchmarks our approach is on average up to 3.7$\times$ and 3$\times$ faster than DS and DASS, respectively.
The performance of our approach is also more stable across varying data distributions.
This is visible in fig. \ref{fig:AreaAndPerfPlot}, where the speedup range bars for DS and DASS dip below 1 more often (which means a \textit{slowdown} over SS).
Our approach is only slower than SS in the \textit{bubbleSort} and \textit{getTanh} benchmarks, and only when the data distribution favors static scheduling.
This is because the frequency of our circuits for those codes is more than 3$\times$ lower than SS due to the critical path overhead of the LSQ.

\begin{table*}[!t]
% \renewcommand{\arraystretch}{1.05}
\setlength{\tabcolsep}{1.8pt}
% if using array.sty, it might be a good idea to tweak the value of
% \extrarowheight as needed to properly center the text within the cells
\caption{Evaluation of our approach against static scheduling (SS), dynamic scheduling \cite{josipovic_dynamatic_2022}, and DASS \cite{dass}. Three sets of benchmarks: 1--4 have control-dependent data dependencies, 5--8 have hazards, and 9--10 have no data-dependent behavior.}
\label{table:benchmarkTable}
\centering
% Some packages, such as MDW tools, offer better commands for making tables
% than the plain LaTeX2e tabular which is used here.
\begin{tabular}{|l | >{\columncolor[gray]{0.9}}r>{\columncolor[gray]{0.9}}r>{\columncolor[gray]{0.9}}r | rrr | >{\columncolor[gray]{0.9}}r>{\columncolor[gray]{0.9}}r>{\columncolor[gray]{0.9}}r>{\columncolor[gray]{0.9}}r | rrrr | >{\columncolor[gray]{0.9}}r>{\columncolor[gray]{0.9}}r>{\columncolor[gray]{0.9}}r>{\columncolor[gray]{0.9}}r |}
\hline
\multirow{2}{*}{} & \multicolumn{3}{>{\columncolor[gray]{0.9}}c|}{\textbf{Area $\times$SS}} & \multicolumn{3}{c|}{\textbf{DSPs $\times$SS}} & \multicolumn{4}{>{\columncolor[gray]{0.9}}c|}{\textbf{FMax (MHz)}} & \multicolumn{4}{c|}{\textbf{Cycles (thousands)}} & \multicolumn{4}{>{\columncolor[gray]{0.9}}c|}{\textbf{Execution Time ($\mu$s)}} \\
% \cline{2-22}
\hhline{~|-|-|-|-|-|-|-|-|-|-|-|-|-|-|-|-|-|-|}
& \multicolumn{1}{>{\columncolor[gray]{0.9}}r}{\cite{josipovic_dynamatic_2022}} & \multicolumn{1}{>{\columncolor[gray]{0.9}}r}{\cite{dass}} & \multicolumn{1}{>{\columncolor[gray]{0.9}}r|}{Us}
& \multicolumn{1}{r}{\cite{josipovic_dynamatic_2022}} & \multicolumn{1}{r}{\cite{dass}} & \multicolumn{1}{r|}{Us}  
& \multicolumn{1}{>{\columncolor[gray]{0.9}}r}{SS} & \multicolumn{1}{>{\columncolor[gray]{0.9}}r}{\cite{josipovic_dynamatic_2022}} & \multicolumn{1}{>{\columncolor[gray]{0.9}}r}{\cite{dass}} & \multicolumn{1}{>{\columncolor[gray]{0.9}}r|}{Us} 
& \multicolumn{1}{r}{SS} & \multicolumn{1}{r}{\cite{josipovic_dynamatic_2022}} & \multicolumn{1}{r}{\cite{dass}} & \multicolumn{1}{r|}{Us}
& \multicolumn{1}{>{\columncolor[gray]{0.9}}r}{SS} & \multicolumn{1}{>{\columncolor[gray]{0.9}}r}{\cite{josipovic_dynamatic_2022}} & \multicolumn{1}{>{\columncolor[gray]{0.9}}r}{\cite{dass}} & \multicolumn{1}{>{\columncolor[gray]{0.9}}r|}{Us} \\

\hline

%%%%%%%%%%%%%%%%%%%%%%%%%%%%%%%%%%%%%%%%  NO LSQ BELOW %%%%%%%%%%%%%%%%%%%%%%%%%%%%%%%%%%%%%%%%
% Intel SS LUTs=115, REG=205,
% Ours LUTs=136, REG=155,
% sparseMatrixPow 
sparseMatrix & 2.5 & 1.3 & 1 & 2 & 1 & 1 & 415 & 161 & 161 & 334 & 1.8--11 & 0.3--11 & 0.3--11 & 0.1--11 & 4.3--26.5 & 1.9--68.3 & 1.9--68.3 & 0.3--32.9 \\

% SS: "alm"  : "168.7", "reg"  : "350",
% Ours 340, 500
% getTanhDouble 
getTanhDouble & 1.1 & 1.2 & 1.5 & 1 & 1 & 1 & 371 & 161 & 161 & 372 & 38 & 1 & 1 & 1  & 102.4 & 6.2 & 6.2 & 2.7 \\

% SS: "alm"  : "723 "reg"  : "2110",
% Ours "alm"  : "1942.5 "reg"  : "4670",
% filterSum 
filterSum & 2.8 & 1.7 & 2 & 1 & 1 & 1 & 425 & 185 & 189 & 411 & 5 & 1--5 & 1--5 & 1--5  & 11.8 & 5.4--27 & 5.3--26.5 & 2.4--12.2 \\

% SS: "alm"  : "930.5 "reg"  : "2435",
% Our:  "alm"  : "2019 "reg"  : "5086",
% vecNormTrans 
vecNormTrans & 2.6 & 2.4 & 2 & 4 & 1.4 & 1 & 374 & 185 & 201 & 379 & 12 & 6.1 & 6.7 & 6.1  & 32.1 & 33 & 33.3 & 16.1 \\

\hline
\textbf{norm. geomean} & \textbf{1.9} & \textbf{1.5} & \textbf{1.4} & \textbf{1.7} & \textbf{1.1} & \textbf{1} & \textbf{1} & \textbf{0.43} & \textbf{0.44} & \textbf{0.94} & \textbf{1} & \textbf{0.14--0.34} & \textbf{0.14--0.34} & \textbf{0.09--0.34} & \textbf{1}  & \textbf{0.33--0.78} & \textbf{0.33--0.78} & \textbf{0.12--0.36} \\
\hline

% Intel SS LUTs=414, REG=548,
% Ours LUTs=750, REG=922,
% histogram 
histogram & 2.1 & 2.4 & 1.2 & 1 & 1 & 1 & 356 & 146 & 146 & 168 & 9    & 1 & 1 & 1 & 25.3 & 6.8 & 6.8 & 6 \\

% SS: "alm"  : "192.0, "reg"  : "346",
% Our:  "alm"  : "376.8, "reg"  : "443",
% bubbleSort
bubbleSort & 7.1 & 7.5 & 1.6 & 1 & 1 & 1 & 447 & 139 & 136 & 168 & 20 & 10--20 & 10--20 & 10--20  & 44.7 & 71.9-143.9 & 73.5-147.1 & 59.5-119 \\

% SS: "alm"  : "169.6 "reg"  : "379",
% Our: 331, 625
% getTanh 
getTanh & 3.5 & 1.3 & 1.7 & 2 & 1 & 1 & 368 & 119 & 119 & 161 & 44-56 & 2.5--66 & 2.5--56 & 1--56 & 120--152 & 21--554.6 & 21--470.6 & 6.2--331.3 \\

% SS: "alm"  : "358.1 "reg"  : "700",
% Ours: 567, 823
% BNN 
BNN & 6.8 & 2.9 & 1.4 & 3 & 1 & 1 & 447 & 124 & 119 & 174 & 60 & 30 & 30 & 10  & 134.2 & 241.9 & 252.1 & 57.5 \\

\hline
\textbf{norm. geomean} & \textbf{4.4} & \textbf{2.8} & \textbf{1.4} & \textbf{1.6} & \textbf{1} & \textbf{1}  & \textbf{1} & \textbf{0.33} & \textbf{0.32} & \textbf{0.42} & \textbf{1} & \textbf{0.2--0.5} & \textbf{0.2--0.32} & \textbf{0.12--0.18} & \textbf{1}  &  \textbf{0.6--1.54} &  \textbf{0.62--1.5} &  \textbf{0.29--0.88} \\
\hline

% covariance
covariance & 3.4 & 1.6 & 1 & 1.8 & 1.8 & 1 & 434 & 86 & 100  & 434 & 68 & 72.9 & 84 & 68  & 156.7 & 847.7 & 840 & 156.7 \\
% gesummv 
gesummv & 1.6 & 1.3 & 1 & 2.2 & 1.7 & 1 & 410 & 113 & 163 & 410 & 65.8 & 262 & 68.8 & 65.8 & 160.5 & 2318.6 & 674.5 & 160.5 \\
\hline
\textbf{norm. geomean} & \textbf{2.3} & \textbf{1.4} & \textbf{1} & \textbf{1.4} & \textbf{1.3} & \textbf{1} &  \textbf{1} & \textbf{0.23} & \textbf{0.3} & \textbf{1} & \textbf{1} & \textbf{2.07} & \textbf{1.13} & \textbf{1} & \textbf{1} & \textbf{8.84} & \textbf{4.75} & \textbf{1} \\
\hline
\hline
\textbf{norm. geomean} & \textbf{2.7} & \textbf{1.8} & \textbf{1.3} & \textbf{1.6} & \textbf{1.1} & \textbf{1} &  \textbf{1} & \textbf{0.3} & \textbf{0.35} & \textbf{0.74} & \textbf{1} & \textbf{0.39-0.71} & \textbf{0.32-0.5} & \textbf{0.22-0.39} & \textbf{1} & \textbf{1.21-2.2} & \textbf{0.99-1.77} & \textbf{0.33-0.68} \\
\hline

\end{tabular}
\end{table*}

\section{Limitations and Future Work} \label{sec:limitations_and_future}

On codes that require an LSQ, the speedup of our implementation over SS is smaller because we suffer the same frequency degradation as DS and DASS.
% The biggest advantage of our approach compared to fully dynamic scheduling is the lower critical paths.
% However, this is only true for codes that don't require a load-store queue (LSQ).
% If required, an LSQ often becomes the critical path in our design, nullifying any frequency advantage over dynamic scheduling.
Thus, a memory disambiguation method with no critical path overhead but with the same throughput as an LSQ is desired for our approach.

Our analysis for marking code for dynamic scheduling could be inaccurate in some cases, because we don't have access to the model of operation latencies used in closed-source back-end compiler.
% While memory latency is easy to obtain, the information about the latency of arithmetic operations is usually part of the compiler back-end in commercial HLS tools and not open source.
In this work, we used a simple model for calculating the recurrence-constrained initiation interval (II), which might, for example, underestimate the extent of operator chaining performed in the compiler back-end.
Ideally, the analysis for finding opportunities for dynamic scheduling should use the same modulo scheduling implementation and latency model as the compiler back-end.

Similarly, in a production compiler, the implementation of our decoupling transformation should not rely on SYCL pipes and kernels, which are user-facing features.
To this end, future work could investigate open-source HLS tools.
A promising development is CIRCT \cite{circt}.
CIRCT is based on the MLIR compiler infrastructure \cite{mlir} and uses different dialects (intermediate representations) to represent hardware with different semantics. 
For example, there exist separate dialects for statically and dynamically scheduled circuits.

\section{Conclusions} \label{sec:conclusion}

We presented an algorithm for identifying code regions amenable to dynamic scheduling in modulo-scheduled HLS and contributed a novel method for realizing dynamically scheduled basic blocks, loops, and out-of-order memory operations in modulo-scheduled HLS.
Our main idea is to decouple parts of control-flow paths that increase the loop initiation interval into separate modulo-scheduled loops connected via latency-insensitive channels.
Our approach is on average 3.7$\times$ faster than a fully dynamically scheduled HLS tool, while using only 1.3$\times$ more area than pure static scheduling.
% and smaller on irregular codes than other HLS scheduling approaches, while not incurring any overheads when dynamic behavior is not needed.
% and a methodology for marking parts of a dataflow circuit for static scheduling.

\section*{Acknowledgment}
This work was partly supported by the UK EPSRC. 
We thank Intel for access to FPGAs through the FPGA DevCloud, and the anonymous reviewers for improving this paper.

\newpage
\bibliographystyle{IEEEtran}
\bibliography{IEEEabrv, references}

\begin{thebibliography}{10}
\providecommand{\url}[1]{#1}
\csname url@samestyle\endcsname
\providecommand{\newblock}{\relax}
\providecommand{\bibinfo}[2]{#2}
\providecommand{\BIBentrySTDinterwordspacing}{\spaceskip=0pt\relax}
\providecommand{\BIBentryALTinterwordstretchfactor}{4}
\providecommand{\BIBentryALTinterwordspacing}{\spaceskip=\fontdimen2\font plus
\BIBentryALTinterwordstretchfactor\fontdimen3\font minus
  \fontdimen4\font\relax}
\providecommand{\BIBforeignlanguage}[2]{{%
\expandafter\ifx\csname l@#1\endcsname\relax
\typeout{** WARNING: IEEEtran.bst: No hyphenation pattern has been}%
\typeout{** loaded for the language `#1'. Using the pattern for}%
\typeout{** the default language instead.}%
\else
\language=\csname l@#1\endcsname
\fi
#2}}
\providecommand{\BIBdecl}{\relax}
\BIBdecl

\bibitem{efficient_communication_analysis_adler}
M.~Pellauer, A.~Parashar, M.~Adler, B.~Ahsan, R.~Allmon, N.~Crago, K.~Fleming,
  M.~Gambhir, A.~Jaleel, T.~Krishna, D.~Lustig, S.~Maresh, V.~Pavlov,
  R.~Rayess, A.~Zhai, and J.~Emer, ``Efficient control and communication
  paradigms for coarse-grained spatial architectures,'' \emph{ACM Trans.
  Comput. Syst.}, 2015.

\bibitem{amazon_f1_fpga}
\BIBentryALTinterwordspacing
``Amazon.com, inc. amazon ec2 f1 instances.'' [Online]. Available:
  \url{https://aws.amazon.com/ec2/instance-types/f1}
\BIBentrySTDinterwordspacing

\bibitem{microsoft_fpga}
A.~M. Caulfield, E.~S. Chung, A.~Putnam, H.~Angepat, J.~Fowers, M.~Haselman,
  S.~Heil, M.~Humphrey, P.~Kaur, J.-Y. Kim, D.~Lo, T.~Massengill, K.~Ovtcharov,
  M.~Papamichael, L.~Woods, S.~Lanka, D.~Chiou, and D.~Burger, ``A cloud-scale
  acceleration architecture,'' in \emph{2016 49th Annual IEEE/ACM International
  Symposium on Microarchitecture (MICRO)}, 2016, pp. 1--13.

\bibitem{modulo_sched}
B.~R. Rau, ``Iterative modulo scheduling: An algorithm for software pipelining
  loops,'' in \emph{Proceedings of the 27th Annual International Symposium on
  Microarchitecture}, 1994.

\bibitem{modulo_sched_canis}
A.~Canis, S.~D. Brown, and J.~H. Anderson, ``Modulo sdc scheduling with
  recurrence minimization in high-level synthesis,'' in \emph{2014 24th
  International Conference on Field Programmable Logic and Applications (FPL)},
  2014, pp. 1--8.

\bibitem{modulo_scheduling_koch}
\BIBentryALTinterwordspacing
J.~Oppermann, A.~Koch, M.~Reuter-Oppermann, and O.~Sinnen, ``Ilp-based modulo
  scheduling for high-level synthesis,'' in \emph{Proceedings of the
  International Conference on Compilers, Architectures and Synthesis for
  Embedded Systems}, ser. CASES '16.\hskip 1em plus 0.5em minus 0.4em\relax New
  York, NY, USA: Association for Computing Machinery, 2016. [Online].
  Available: \url{https://doi.org/10.1145/2968455.2968512}
\BIBentrySTDinterwordspacing

\bibitem{Arvind_Culler_1986}
Arvind and D.~E. Culler, ``Dataflow architectures,'' \emph{Annual Review of
  Computer Science}, 1986.

\bibitem{dynamatic_ltl}
\BIBentryALTinterwordspacing
J.~Xu, E.~Murphy, J.~Cortadella, and L.~Josipovic, ``Eliminating excessive
  dynamism of dataflow circuits using model checking,'' in \emph{Proceedings of
  the 2023 ACM/SIGDA International Symposium on Field Programmable Gate
  Arrays}, ser. FPGA '23.\hskip 1em plus 0.5em minus 0.4em\relax New York, NY,
  USA: Association for Computing Machinery, 2023, p. 27–37. [Online].
  Available: \url{https://doi.org/10.1145/3543622.3573196}
\BIBentrySTDinterwordspacing

\bibitem{josipovic_dynamatic_2022}
L.~Josipović, A.~Guerrieri, and P.~Ienne, ``From c/c++ code to
  high-performance dataflow circuits,'' \emph{IEEE Transactions on
  Computer-Aided Design of Integrated Circuits and Systems}, 2022.

\bibitem{dass}
J.~Cheng, L.~Josipović, G.~A. Constantinides, P.~Ienne, and J.~Wickerson,
  ``Dass: Combining dynamic amp; static scheduling in high-level synthesis,''
  \emph{IEEE Transactions on Computer-Aided Design of Integrated Circuits and
  Systems}, 2022.

\bibitem{Carloni_McMillan_Sangiovanni_Vincentelli_2001}
L.~Carloni, K.~McMillan, and A.~Sangiovanni-Vincentelli, ``Theory of
  latency-insensitive design,'' \emph{IEEE Transactions on Computer-Aided
  Design of Integrated Circuits and Systems}, 2001.

\bibitem{Cortadella_Kishinevsky_Grundmann}
J.~Cortadella, M.~Kishinevsky, and B.~Grundmann, ``Synthesis of synchronous
  elastic architectures,'' in \emph{2006 43rd ACM/IEEE Design Automation
  Conference}, 2006.

\bibitem{Kam_Kishinevsky_Cortadella_micro_pipelining}
T.~Kam, M.~Kishinevsky, J.~Cortadella, and M.~Galceran-Oms,
  ``Correct-by-construction microarchitectural pipelining,'' in \emph{2008
  IEEE/ACM International Conference on Computer-Aided Design}, 2008, pp.
  434--441.

\bibitem{huang2013elastic}
Y.~Huang, P.~Ienne, O.~Temam, Y.~Chen, and C.~Wu, ``Elastic cgras,'' in
  \emph{Proceedings of the ACM/SIGDA International Symposium on Field
  Programmable Gate Arrays}, ser. FPGA '13, 2013.

\bibitem{elastic_flow}
M.~Tan, G.~Liu, R.~Zhao, S.~Dai, and Z.~Zhang, ``Elasticflow: A
  complexity-effective approach for pipelining irregular loop nests,'' in
  \emph{2015 IEEE/ACM International Conference on Computer-Aided Design
  (ICCAD)}, 2015, pp. 78--85.

\bibitem{Townsend_Kim_Edwards_2017}
R.~Townsend, M.~A. Kim, and S.~A. Edwards, ``From functional programs to
  pipelined dataflow circuits,'' in \emph{Proceedings of the 26th International
  Conference on Compiler Construction}, 2017.

\bibitem{sycl_khronos}
\BIBentryALTinterwordspacing
``Khronos sycl registry.'' [Online]. Available:
  \url{https://registry.khronos.org/SYCL/}
\BIBentrySTDinterwordspacing

\bibitem{vitis}
\BIBentryALTinterwordspacing
``Amd xilinx vitis.'' [Online]. Available:
  \url{https://www.xilinx.com/products/design-tools/vitis.html}
\BIBentrySTDinterwordspacing

\bibitem{Edwards2019CompositionalDC}
S.~A. Edwards, R.~Townsend, M.~Barker, and M.~A. Kim, ``Compositional dataflow
  circuits,'' \emph{ACM Transactions on Embedded Computing Systems (TECS)},
  vol.~18, 2019.

\bibitem{Venkataramani2001}
G.~Venkataramani, M.~Budiu, T.~Chelcea, and S.~C. Goldstein, ``{C to
  Asynchronous Dataflow Circuits: An End-to-End Toolflow},'' \emph{IWLS'01},
  2001.

\bibitem{Rosen1988GlobalVN}
B.~K. Rosen, M.~N. Wegman, and F.~K. Zadeck, ``Global value numbers and
  redundant computations,'' in \emph{ACM-SIGACT Symposium on Principles of
  Programming Languages}, 1988.

\bibitem{dynamatic_gated_ssa}
A.~Elakhras, A.~Guerrieri, L.~Josipović, and P.~Ienne, ``Unleashing
  parallelism in elastic circuits with faster token delivery,'' in \emph{2022
  32nd International Conference on Field-Programmable Logic and Applications
  (FPL)}, 2022, pp. 253--261.

\bibitem{cheng_finding_static}
\BIBentryALTinterwordspacing
J.~Cheng, J.~Wickerson, and G.~A. Constantinides, ``Finding and finessing
  static islands in dynamically scheduled circuits,'' in \emph{Proceedings of
  the 2022 ACM/SIGDA International Symposium on Field-Programmable Gate
  Arrays}, ser. FPGA '22.\hskip 1em plus 0.5em minus 0.4em\relax New York, NY,
  USA: Association for Computing Machinery, 2022, p. 89–100. [Online].
  Available: \url{https://doi.org/10.1145/3490422.3502362}
\BIBentrySTDinterwordspacing

\bibitem{dswp}
G.~Ottoni, R.~Rangan, A.~Stoler, and D.~August, ``Automatic thread extraction
  with decoupled software pipelining,'' in \emph{38th Annual IEEE/ACM
  International Symposium on Microarchitecture (MICRO'05)}, 2005, pp. 12
  pp.--118.

\bibitem{decoupled_memory_wawrzynek}
S.~Cheng and J.~Wawrzynek, ``Architectural synthesis of computational pipelines
  with decoupled memory access,'' in \emph{2014 International Conference on
  Field-Programmable Technology (FPT)}, 2014, pp. 83--90.

\bibitem{opencl_moc}
\BIBentryALTinterwordspacing
N.~Kapre and H.~Patel, ``Applying models of computation to opencl pipes for
  fpga computing,'' in \emph{Proceedings of the 5th International Workshop on
  OpenCL}, ser. IWOCL 2017.\hskip 1em plus 0.5em minus 0.4em\relax New York,
  NY, USA: Association for Computing Machinery, 2017. [Online]. Available:
  \url{https://doi.org/10.1145/3078155.3078163}
\BIBentrySTDinterwordspacing

\bibitem{Hauck_DeHon_2008}
S.~Hauck and A.~DeHon, \emph{Reconfigurable Computing: The Theory and Practice
  of FPGA-Based Computation}.\hskip 1em plus 0.5em minus 0.4em\relax San
  Francisco, CA, USA: Morgan Kaufmann Publishers Inc., 2007.

\bibitem{mlir}
C.~Lattner, M.~Amini, U.~Bondhugula, A.~Cohen, A.~Davis, J.~Pienaar, R.~Riddle,
  T.~Shpeisman, N.~Vasilache, and O.~Zinenko, ``{{MLIR}}: Scaling compiler
  infrastructure for domain specific computation,'' in \emph{2021 {{IEEE/ACM}}
  International Symposium on Code Generation and Optimization (CGO)}, 2021, pp.
  2--14.

\bibitem{sycl_kernel_fusion}
\BIBentryALTinterwordspacing
V.~P\'{e}rez, L.~Sommer, V.~Lom\"{u}ller, K.~Narasimhan, and M.~Goli,
  ``User-driven online kernel fusion for sycl,'' \emph{ACM Trans. Archit. Code
  Optim.}, vol.~20, no.~2, mar 2023. [Online]. Available:
  \url{https://doi.org/10.1145/3571284}
\BIBentrySTDinterwordspacing

\bibitem{Ferrante1987ThePD}
J.~Ferrante, K.~J. Ottenstein, and J.~D. Warren, ``The program dependence graph
  and its use in optimization,'' in \emph{TOPL}, 1987.

\bibitem{dependence_distance_hls}
J.~Cheng, J.~Wickerson, and G.~A. Constantinides, ``Exploiting the correlation
  between dependence distance and latency in loop pipelining for hls,'' in
  \emph{2021 31st International Conference on Field-Programmable Logic and
  Applications (FPL)}, 2021, pp. 341--346.

\bibitem{poly_hls}
A.~Morvan, S.~Derrien, and P.~Quinton, ``Efficient nested loop pipelining in
  high level synthesis using polyhedral bubble insertion,'' in \emph{2011
  International Conference on Field-Programmable Technology}, 2011.

\bibitem{poly_cc}
M.-W. Benabderrahmane, L.-N. Pouchet, A.~Cohen, and C.~Bastoul, ``The
  polyhedral model is more widely applicable than you think,'' in
  \emph{Compiler Construction}, R.~Gupta, Ed.\hskip 1em plus 0.5em minus
  0.4em\relax Berlin, Heidelberg: Springer Berlin Heidelberg, 2010, pp.
  283--303.

\bibitem{dynamic_inter_block_cheng}
J.~Cheng, L.~Josipović, G.~A. Constantinides, and J.~Wickerson, ``Dynamic
  inter-block scheduling for hls,'' in \emph{2022 32nd International Conference
  on Field-Programmable Logic and Applications (FPL)}, 2022, pp. 243--252.

\bibitem{online_poy_hls}
J.~Liu, S.~Bayliss, and G.~A. Constantinides, ``Offline synthesis of online
  dependence testing: Parametric loop pipelining for hls,'' in \emph{2015 IEEE
  23rd Annual International Symposium on Field-Programmable Custom Computing
  Machines}, 2015, pp. 159--162.

\bibitem{Josipovic_Brisk_Ienne_2017}
L.~Josipovic, P.~Brisk, and P.~Ienne, ``An out-of-order load-store queue for
  spatial computing,'' \emph{ACM Transactions on Embedded Computing Systems},
  2017.

\bibitem{decoupled_access_exec}
J.~E. Smith, ``Decoupled access/execute computer architectures,'' in
  \emph{Proceedings of the 9th Annual Symposium on Computer Architecture}, ser.
  ISCA '82.\hskip 1em plus 0.5em minus 0.4em\relax Washington, DC, USA: IEEE
  Computer Society Press, 1982, p. 112–119.

\bibitem{decoupled_memory_prefetching}
T.~Chen and G.~E. Suh, ``Efficient data supply for hardware accelerators with
  prefetching and access/execute decoupling,'' in \emph{2016 49th Annual
  IEEE/ACM International Symposium on Microarchitecture (MICRO)}, 2016, pp.
  1--12.

\bibitem{Thielmann_Load_Speculation}
B.~Thielmann, J.~Huthmann, and A.~Koch, ``Memory latency hiding by load value
  speculation for reconfigurable computers,'' \emph{ACM Trans. Reconfigurable
  Technol. Syst.}, 2012.

\bibitem{straight_to_the_queue}
\BIBentryALTinterwordspacing
A.~Elakhras, R.~Sawhney, A.~Guerrieri, L.~Josipovic, and P.~Ienne, ``Straight
  to the queue: Fast load-store queue allocation in dataflow circuits,'' in
  \emph{Proceedings of the 2023 ACM/SIGDA International Symposium on Field
  Programmable Gate Arrays}, ser. FPGA '23.\hskip 1em plus 0.5em minus
  0.4em\relax New York, NY, USA: Association for Computing Machinery, 2023, p.
  39–45. [Online]. Available: \url{https://doi.org/10.1145/3543622.3573050}
\BIBentrySTDinterwordspacing

\bibitem{LLVM_CGO04}
C.~Lattner and V.~Adve, ``{LLVM}: A compilation framework for lifelong program
  analysis and transformation,'' in \emph{CGO}, 2004.

\bibitem{benchmarks_jianyi_cheng_2019}
\BIBentryALTinterwordspacing
J.~Cheng, ``{JianyiCheng: HLS\_Benchmarks\_First\_Release},'' Dec. 2019.
  [Online]. Available: \url{https://doi.org/10.5281/zenodo.3561115}
\BIBentrySTDinterwordspacing

\bibitem{mem_disamb_approaches}
H.~Wong, V.~Betz, and J.~Rose, ``Efficient methods for out-of-order load/store
  execution for high-performance soft processors,'' in \emph{2013 International
  Conference on Field-Programmable Technology (FPT)}, 2013, pp. 442--445.

\bibitem{circt}
\BIBentryALTinterwordspacing
 [Online]. Available: \url{https://circt.llvm.org/}
\BIBentrySTDinterwordspacing

\end{thebibliography}

\end{document}